\documentclass[msom,sglanonrev]{informs4_plain}

\usepackage{eqndefns-left} 
\RequirePackage{tgtermes}
\RequirePackage{newtxtext}
\RequirePackage{newtxmath}
\RequirePackage{bm}
\RequirePackage{endnotes}

\OneAndAHalfSpacedXII 

\usepackage{natbib}
 \bibpunct[, ]{(}{)}{,}{a}{}{,}%
 \def\bibfont{\small}%
\usepackage{caption}
\usepackage[labelfont=sf]{subcaption}
\TheoremsNumberedThrough     
\ECRepeatTheorems

\EquationsNumberedThrough    

\MANUSCRIPTNO{}

\begin{document}

\RUNAUTHOR{ }

\RUNTITLE{Managing Cognitive Bias in Human Labeling Operations}

\TITLE{Managing Cognitive Bias in Human Labeling Operations for Rare-Event AI: Evidence from a Field Experiment}

\ARTICLEAUTHORS{%
\AUTHOR{Gunnar P. Epping}
\AFF{Cognitive Science Program, Indiana University, Indiana, USA; Department of Psychological and Brain Sciences, Indiana University, Indiana, USA; Centaur Labs, Massachusetts, USA, \EMAIL{gunnarepping@gmail.com}}

\AUTHOR{Andrew Caplin}
\AFF{Department of Economics, New York University, New York, USA, \EMAIL{ac1@nyu.edu}}

\AUTHOR{Erik Duhaime}
\AFF{Centaur Labs, Massachusetts, USA, \EMAIL{erik@centaurlabs.com}}

\AUTHOR{William R. Holmes}
\AFF{Cognitive Science Program, Indiana University, Indiana, USA; Department of Mathematics, Indiana University, Indiana, USA, \EMAIL{wrholmes@iu.edu}}

\AUTHOR{Daniel Martin}
\AFF{Department of Economics, University of California, Santa Barbara, USA, \EMAIL{danielmartin@ucsb.edu}}

\AUTHOR{Jennifer S. Trueblood}
\AFF{Cognitive Science Program, Indiana University, Indiana, USA; Department of Psychological and Brain Sciences, Indiana University, Indiana, USA, \EMAIL{jstruebl@iu.edu}}
} 

\ABSTRACT{%
\textit{\textbf{Problem definition:}} Many operational AI systems depend on large-scale human annotation to detect rare but consequential events (e.g., fraud, defects, and medical abnormalities). When positives are rare, the prevalence effect induces systematic cognitive biases that inflate misses and can propagate through the AI lifecycle via biased training labels. \textit{\textbf{Methodology/results:}} We analyze prior experimental evidence and run a field experiment on DiagnosUs, a medical crowdsourcing platform, in which we hold the true prevalence in the unlabeled stream fixed (20\% blasts) while varying (i) the prevalence of positives in the gold-standard feedback stream (20\% vs. 50\%) and (ii) the response interface (binary labels vs. elicited probabilities). We then post-process probabilistic labels using a linear-in-log-odds recalibration approach at the worker and crowd levels, and train convolutional neural networks on the resulting labels. Balanced feedback and probabilistic elicitation reduce rare-event misses, and pipeline-level recalibration substantially improves both classification performance and probabilistic calibration; these gains carry through to downstream CNN reliability out of sample. \textit{\textbf{Managerial implications:}} Labeling operations can mitigate systematic under-detection of rare positives by tuning feedback prevalence, eliciting probabilistic judgments, and adding a lightweight recalibration step prior to aggregation and model training.
}%



\maketitle

\section{Introduction}
\label{sec:intro}

Artificial intelligence (AI) is increasingly embedded in operational decision making.
Organizations now deploy machine learning (ML) systems to detect rare but consequential events such as fraud, defects, safety incidents, abnormal medical findings, and content policy violations.
In these settings, model performance is often limited not by model architecture, but by the \emph{AI lifecycle operations} that produce, maintain, and monitor high-quality training and evaluation data.
A central bottleneck is data annotation \citep{orting2019survey, willemink2020preparing}: many firms rely on large pools of human annotators to label items, often with built-in quality assurance (QA) mechanisms such as inter-rater redundancy, majority vote aggregation, and interleaved gold-standard (GS) items that provide feedback and score workers.

A key challenge is that rare-event environments systematically distort human judgment.
When the base rate of the positive class is low (e.g., blast blood cells are rare), people tend to respond ``negative'' more often, producing an increase in \emph{misses} (false negatives) with a corresponding increase in \emph{false alarms} (false positives); when the positive class is common, the opposite pattern emerges.
This phenomenon, the \emph{prevalence effect}, has been documented in visual search (e.g. radiology tasks and airport baggage screening) and perceptual decision making (e.g., medical image classification) \citep{wolfe2005rare,wolfe2007low,gur2003prevalence,horowitz2017prevalence,trueblood2021disentangling}.
Prior work attributes the effect to systematic cognitive biases in the way people evaluate and respond to information \citep{green1966signal,wolfe2010varying,trueblood2021disentangling}, implying that the effect is not due to noise that can be averaged out.
For operations, this is particularly problematic because misses are often costly (e.g., missed fraud, missed defects, missed abnormalities), and because many AI pipelines implicitly assume that additional labels and simple aggregation will reliably improve data quality.

Ideally, redundancy and aggregation would cancel individual errors, yielding more accurate labels for downstream AI. However, if many annotators are exposed to a similar base rate and a similar task environment, their errors can become positively correlated due to prevalence-induced biases.
In that case, the standard wisdom of the crowd (WoC) remedy may provide only limited protection and can even fail in extreme conditions \citep{hong2004groups,davis2014crowd}.
Additionally, biased labels can propagate through the AI lifecycle.
Training labels shape fitted models, model calibration, and downstream operational decisions; thus, a systematic labeling bias can become a systematic model bias.
This risk is amplified by the fact that data labeling pipelines typically include operational choices that are \emph{not} visible to annotators, such as how often GS items appear, which items are used for feedback, and how performance is scored.
These choices can change the \emph{experienced} prevalence during annotation even when the \emph{true} prevalence in production data remains fixed.

This paper studies a practical AI-in-operations problem: \emph{How should an organization design the labeling and feedback process for rare-event classification so that aggregated labels, and models trained on those labels, remain accurate and well-calibrated?}
We focus on three operational levers that are commonly available in real annotation workflows:
(i) the prevalence of positives in the stream of feedback/GS items (a training and QA design choice),
(ii) the response interface (binary labels versus probabilistic beliefs), and
(iii) post-processing of labels via recalibration.
Our goal is not to argue that prevalence effects exist, but to provide causal evidence on pipeline design choices that mitigate prevalence-induced distortions and to quantify the downstream consequences for trained ML models.

\paragraph{Overview and research questions.}
We investigate three questions that are central to managing AI lifecycle operations under class imbalance.
First, holding the prevalence in the unlabeled (production) stream fixed, does changing the prevalence in the GS feedback stream shift human decision behavior and alter the resulting labels?
Second, can a richer interface that elicits subjective probabilities reduce rare-event misses relative to binary labeling, without requiring expert annotators?
Third, can a platform correct remaining systematic biases through a scalable recalibration approach, and do such corrections improve not only label quality but also the reliability of ML models trained on these labels?

\paragraph{Empirical setting and approach.}
We study these questions in the context of medical image annotation, where abnormalities are typically rare and misses can be costly.
Across both studies, the task is to classify white blood cell images as \emph{blast} versus \emph{non-blast} (a rare-event detection problem in practice).
Study~1 analyzes existing experimental data from \citet{trueblood2021disentangling} to establish that prevalence-induced biases observed at the individual level can persist at the aggregate level and that naive aggregation can fail in extreme prevalence regimes, clarifying when additional redundancy does not reliably improve performance.

Study~2 provides causal field evidence in a real-world annotation operation.
We run a field experiment on DiagnosUs, a crowdsourcing platform for medical and scientific data annotation \citep{press2021centaur}.
In DiagnosUs contests, unlabeled QA images are interleaved with GS images that provide feedback and determine leaderboard rewards.
This structure mirrors many operational labeling pipelines in which worker training and evaluation are intertwined with production labeling.
We exploit this structure to manipulate an algorithmic design parameter: the prevalence of positive cases in the GS stream.
In all conditions, the unlabeled QA stream has low prevalence (20\% blasts), reflecting the rarity of abnormalities; we vary whether the GS stream matches this prevalence (20\%) or is balanced (50\%).
Crossing this lever with the response interface yields four experimental conditions: binary classification versus elicited beliefs (subjective probabilities), each under unbalanced (20\%) or balanced (50\%) GS prevalence.

To address residual systematic bias, we evaluate a post-processing intervention based on recalibration of probabilistic judgments.
When elicited beliefs are available, systematic under- or overprediction manifests as miscalibration and can be corrected using the linear-in-log-odds (LLO) transformation \citep{birnbaum1996violations,turner2014forecast,han2022recalibrating,epping2025harnessing}.
We implement LLO both at the \emph{individual} level (using each worker's GS trials as a calibration set) and at the \emph{crowd} level (using aggregated GS labels as a calibration set), allowing us to compare worker-level versus pipeline-level correction.

Finally, to connect labeling operations to downstream AI performance, we train convolutional neural networks (CNNs) on the resulting label variants and evaluate the models against ground truth.
This step quantifies how operational labeling design choices translate into changes in model error tradeoffs (misses versus false alarms) and probabilistic reliability (calibration), which are critical for operational deployment. Calibration is important because many human-AI collaborative tasks rely on model predicted probabilities. When probabilities are miscalibrated, end-users may over- or under-react to model outputs even when overall model accuracy is high. We measure calibration using expected calibration error (ECE) \citep{guo2017calibration} alongside miss and false alarm rates, aligning evaluation with operational concerns about threshold-based decision making under asymmetric costs.

\paragraph{Preview of main findings.}
We find that the GS feedback policy materially shapes labeling outcomes even when the true prevalence in the unlabeled stream is held fixed.
Balanced GS feedback reduces the extreme miss-heavy error profile typical of low-prevalence environments, producing a more balanced miss/false-alarm tradeoff.
Eliciting probabilistic beliefs provides additional information that improves aggregation relative to binary labels in low-prevalence conditions.
Most importantly, pipeline-level recalibration of aggregated probabilities substantially improves both error rates and calibration, mitigating systematic underestimation of rare positives that persists after naive aggregation.
These improvements carry through to downstream AI: CNNs trained on recalibrated crowd labels exhibit better alignment between predicted probabilities and ground truth and improved rare-event detection performance out of sample.

\paragraph{Contributions to AI in Operations.}
This work contributes to the emerging operations literature on AI lifecycle management in three ways.
First, we identify prevalence-induced biases as an operational failure in human labeling pipelines and show when redundancy and simple aggregation are insufficient.
Second, we provide causal field evidence on a concrete, implementable design lever, the composition of the GS feedback stream, that many platforms and organizations can adjust without changing the underlying production distribution.
Third, we evaluate a scalable post-processing policy (individual and crowd-level recalibration) that can be layered onto existing pipelines to improve probabilistic reliability, and we quantify downstream effects on trained ML models, connecting data operations choices to model behavior.

\paragraph{Organization.}
The remainder of the paper is organized as follows.
Section~\ref{sec:study1} presents Study~1 and documents how prevalence-induced biases propagate from individuals to aggregated labels.
Section~\ref{sec:study2} presents the DiagnosUs field experiment, including design, incentive structure, and labeling results under alternative pipeline policies.
Section~\ref{sec:ml} evaluates downstream CNNs trained on the resulting label variants.
Section~\ref{sec:discussion} discusses implications for the design and governance of rare-event AI pipelines and outlines directions for future work. Section~\ref{sec:conclusion} briefly concludes.

\section{Study 1}
\label{sec:study1}
The data examined in Study~1 were also reported by \cite{trueblood2021disentangling}.
Study~1 was composed of two experiments: Study~1a and Study~1b.
The previous work used data from Studies~1a and~1b to identify and dissociate cognitive biases due to extreme prevalence environments.
The present work uses data from Studies~1a and~1b to show that the prevalence effect at the individual level can spread to the crowd level. We also evaluate how the prevalence effect at the crowd level is influenced by the crowd size.
The following section provides an overview of the methods for the two studies, and additional details regarding the study methods are available in \cite{trueblood2021disentangling}. The data is available on the Open Science Framework at \url{osf.io/4n7sr}.

\subsection{Behavioral Methods}

\subsubsection{Participants.}
39 Vanderbilt University undergraduate students participated in Study~1a for course credit.
The mean age of the 39 students was 19.7 and 77\% were female.
57 Vanderbilt University undergraduate students participated in Study~1b for course credit.
The mean age of the 57 students was 19.2 and 58\% were female.

\subsubsection{Materials.}
\begin{figure}
\FIGURE
{%
\centering
\subcaptionbox{Blast cell.\label{fig:blast-cell}}[.45\textwidth]{\includegraphics[width=\linewidth]{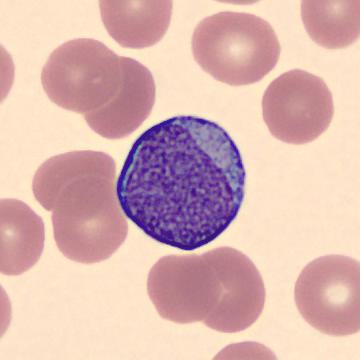}}
\hfill
\subcaptionbox{non-blast cell.\label{fig:non-blast-cell}}[.45\textwidth]{\includegraphics[width=\linewidth]{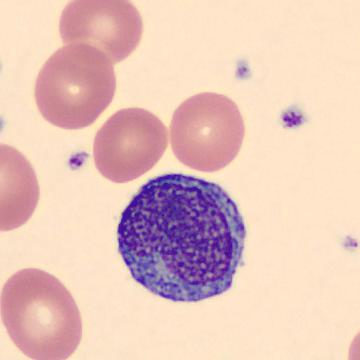}}
}
{Examples of white blood cell stimuli used in Studies~1 and~2.\label{fig:cells}}
{}
\end{figure}

The same set of stimuli were used in both Study~1a and Study~1b.
The stimuli were 300 digital images of Wright-stained white blood cells taken from anonymized patient peripheral blood smears at Vanderbilt University Medical Center (VUMC) and were supplied with ground-truth labels by three hematopathology faculty from the Department of Pathology, Microbiology and Immunology at VUMC \citep{trueblood2018impact}.
Half of the white blood cell images were blast cells and the other half were non-blast cells.
Examples of both blast and non-blast cells can be seen in Figure~\ref{fig:cells}.

\subsubsection{Procedure.}
In both studies, participants were first presented with instructions informing them that they would be classifying white blood cells as either ``blast'' (cancerous) or ``non-blast'' (non-cancerous) cells.
After reviewing the instructions for the task and going through several practice blocks where they received trial-by-trial feedback, the participants began the main portion of the experiment.
Participants did not receive any feedback during the main portion of the experiment.

In Study~1a, the main portion of the experiment consisted of 21 blocks with 48 trials per block.
There were three different block types which differed based on the prevalence rate of blast images in each block: high (75\% blast), equal (50\% blast), and low (25\% blast) prevalence.
Participants were informed of the block type (the proportion of blast images) before beginning each block.
Each participant went through seven blocks of each type, resulting in the 21 blocks in the main portion of the experiment. The order of the blocks was randomized.

In Study~1b, the main portion of the experiment consisted of 14 blocks with 80 trials per block.
After reviewing the instructions for the task and going through several practice blocks, the participants were randomly assigned to one of two groups (high prevalence group and low prevalence group) for the main portion of the experiment.
For the high prevalence group, there were two different block types: high (90\% blast) and equal (50\% blast) prevalence.
For the low prevalence group, there were two different block types: low (10\% blast) and equal (50\% blast) prevalence.
Participants were informed of the block type (the proportion of blast images) before beginning each block.
At the beginning of the main portion of the experiment, both groups completed two equal prevalence blocks.
After that, the high prevalence group completed 12 high prevalence blocks and the low prevalence group completed 12 low prevalence blocks.

\subsection{Modeling Methods}

\subsubsection{Inclusion Criteria.}
Participant responses were filtered such that they could contribute at most one response per image per prevalence block type.
In the event a participant saw an image more than once within a given prevalence block type, we only included their first response in the data analysis.
For example, if a given participant in the high prevalence group in Study~1b saw the same image during the first test block (50\% prevalence), the third test block (90\% prevalence), and the sixth test block (90\% prevalence), only their responses for this image on the first and third test blocks were included in further analysis.
We allowed participants to contribute multiple judgments per image across the different prevalence rates to ensure we recorded several judgments for each image in the low prevalence conditions, but we restricted participants to only contribute one judgment per image within each prevalence block type to minimize the impact of memory on participants' decision-making processes.

\subsubsection{WoC Classification.}
A WoC classification for a given image was generated by randomly sampling seven individual judgments for that image and taking the majority class (either ``blast'' or ``non-blast'', depending on which received more votes). WoC classifications were generated independently for the three different prevalence block types in Study~1a and Study~1b. This resulted in six different sets of WoC classifications for Study~1. We chose to generate each WoC classification using only seven individual judgments because some of the images in the low/high prevalence conditions only received seven judgments. We also note that in real-world crowdsourcing tasks, the aim is to strike a balance between maximizing accuracy and minimizing cost. Given that it costs money to collect more individual judgments per image and gains in accuracy decrease as the number of judgments increases \citep{budescu2015identifying}, real-world medical data annotation tasks rarely collect more than 10 annotations per image \citep{duhaime2023nonexpert}. The process of randomly sampling seven judgments and taking the majority judgment as the WoC classification was repeated 100 times for each image across all prevalence block types, resulting in 100 WoC datasets for each of the six prevalence block types in Study~1.

\subsection{Results}

\subsubsection{Individual Results.}

\begin{figure}
\FIGURE
{%
\centering
\begin{minipage}{.8\textwidth}
\centering
\subcaptionbox{Study~1a miss rate and false alarm rate\label{fig:study1a-indiv-mr-far}}{\includegraphics[width=\linewidth]{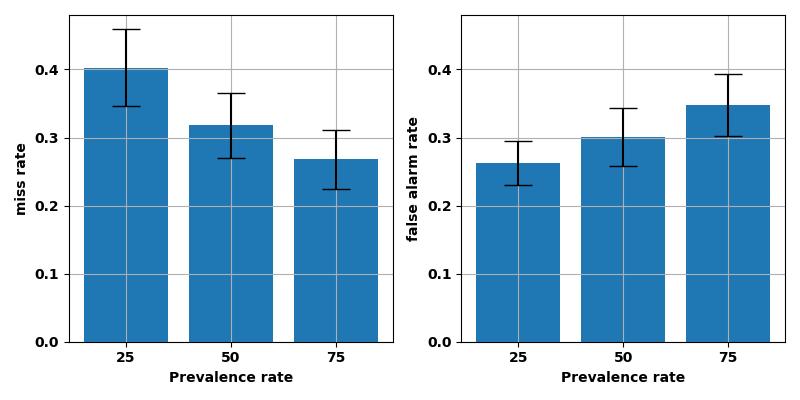}}\\[1.0ex]
\subcaptionbox{Study~1b miss rate and false alarm rate\label{fig:study1b-indiv-mr-far}}{\includegraphics[width=\linewidth]{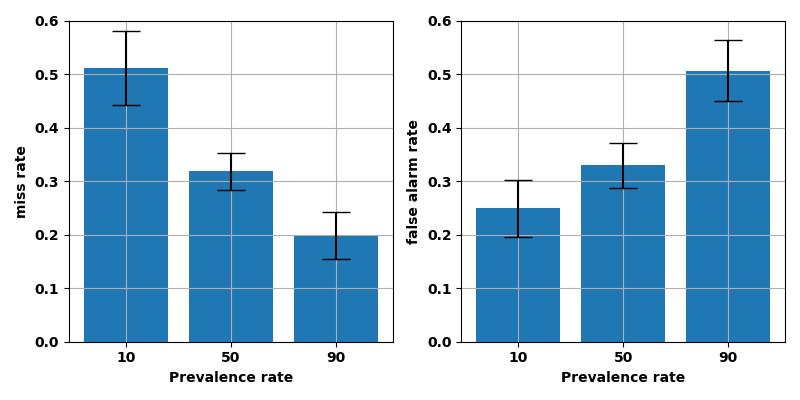}}
\end{minipage}
}
{Study~1 individual-level miss rates and false alarm rates, calculated separately for each prevalence rate.\label{fig:study1-indiv-mr-far}}
{}
\end{figure}

We found in both Studies~1a and~1b that the individual-level miss rates decreased as the prevalence rate increased and the individual-level false alarm rates increased as the prevalence rate increased, which is illustrated in Figure~\ref{fig:study1-indiv-mr-far}.
This aligns with the classical prevalence effect observed in \citet{horowitz2017prevalence} and \citet{wolfe2007low}.

\subsubsection{Crowd Results.}

\begin{figure}
\FIGURE
{%
\centering
\begin{minipage}{.8\textwidth}
\centering
\subcaptionbox{Study~1a miss rate and false alarm rate\label{fig:study1a-crowd-mr-far}}{\includegraphics[width=\linewidth]{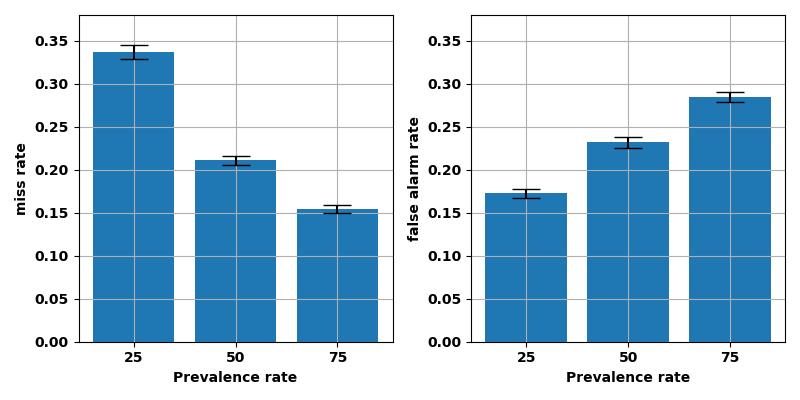}}\\[1.0ex]
\subcaptionbox{Study~1b miss rate and false alarm rate\label{fig:study1b-crowd-mr-far}}{\includegraphics[width=\linewidth]{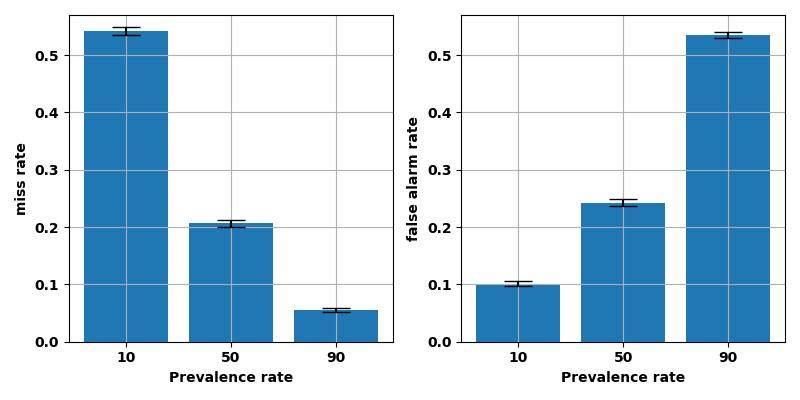}}
\end{minipage}
}
{Study~1 crowd-level miss rates and false alarm rates, calculated separately for each prevalence rate.\label{fig:study1-crowd-mr-far}}
{}
\end{figure}

The prevalence effect observed at the individual level is clearly reflected at the crowd level as well in Studies~1a and~1b, displayed in Figure~\ref{fig:study1-crowd-mr-far}.
That is, miss rates decreased and false alarm rates increased as the prevalence rate increased.

For the most part, the miss rate and false alarm rate in each condition are lower at the crowd level compared to the individual level, suggesting that the WoC applies despite the fact that the prevalence effect biases annotators similarly.
That is, the consensus opinion from a group of annotators is more likely to be correct compared to an opinion from a randomly selected individual within the group.

The two exceptions to this are the miss rate in the 10\% prevalence block type and the false alarm rate in the 90\% prevalence block type.
In these two cases, the error rates appear to be higher at the crowd level compared to the individual level.
This finding was anticipated because, focusing on the first case, the mean individual-level miss rate in the 10\% prevalence block type is 51.1\%.
To illustrate why we anticipated this, suppose the likelihood of a random individual being correct is $p$ and all members of the crowd have the same likelihood of being correct.
With a crowd size of one, the likelihood of the crowd being correct is then equal to $p$.
With a crowd size of three, the likelihood of being correct is equal to the likelihood of at least two of the three judgments being correct which is equal to $p^3 + 3p^2(1-p)$.
Therefore, the likelihood of the crowd being correct with a crowd size of one is greater than the likelihood of the crowd being correct with a crowd size of three when $p > p^3 + 3p^2(1-p)$.
Solving for $p$, one finds that the inequality is true when $p < 0.5$.
While this example is a special case, it illustrates how the crowd can be less accurate than a randomly selected individual when the randomly selected individual is more likely to be incorrect, which is the situation we see in the two extreme prevalence blocks here.

\begin{figure}
\FIGURE
{%
\centering
\begin{minipage}{.8\textwidth}
\centering
\subcaptionbox{Study~1a crowd-level miss rate and false alarm rate by crowd size\label{fig:study1a-crowd-size}}{\includegraphics[width=\linewidth]{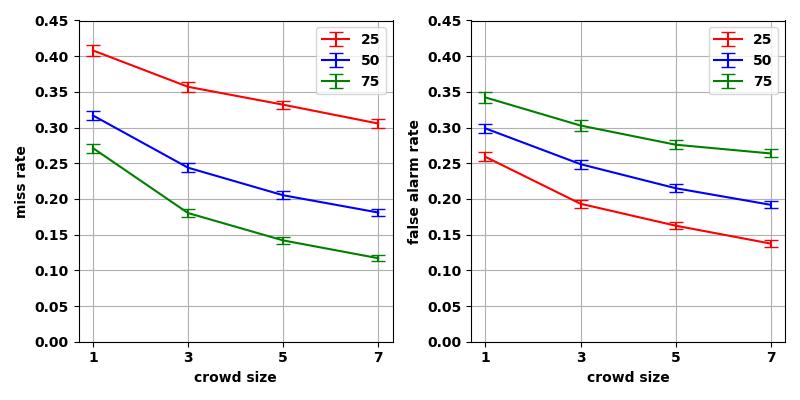}}\\[1.0ex]
\subcaptionbox{Study~1b crowd-level miss rate and false alarm rate as a function of crowd size\label{fig:study1b-crowd-size}}{\includegraphics[width=\linewidth]{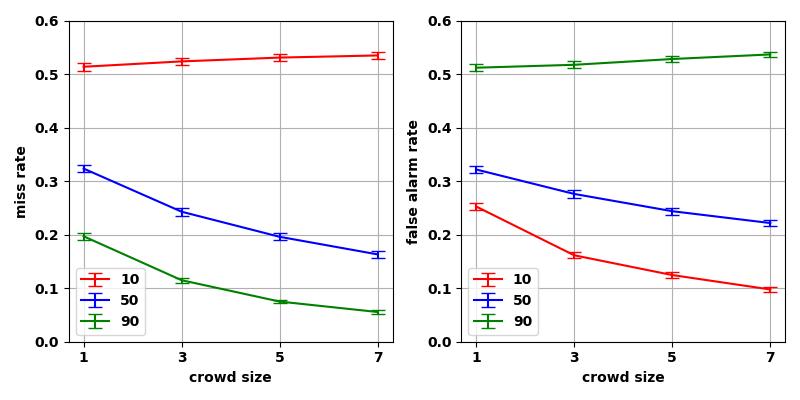}}
\end{minipage}
}
{Study~1 crowd-level miss rates and false alarm rate by crowd size.\label{fig:study1-crowd-size}}
{}
\end{figure}

To investigate how crowd size impacts the prevalence effect, we repeated the process of creating the WoC datasets with one, three, five, and seven judgments per image and plotted the results in Figure~\ref{fig:study1-crowd-size}.
We find that the miss rate and false alarm rate across almost all conditions decrease as the number of judgments increases, except for the miss rate in the Study~1b 10\% prevalence block type and the false alarm rate in the Study~1b 90\% prevalence block type.
These two error rates increase because, at the individual level, the miss rate in Study~1b 10\% prevalence condition and the false alarm rate in Study~1b 90\% prevalence condition are already above 50\%.
Observing the error rates increase as the number of judgments increases for the two extreme prevalence blocks aligns with the special case introduced earlier demonstrating that when the average individual from a crowd is usually incorrect, the likelihood of the crowd also being incorrect increases with crowd size.

\subsection{Discussion}
Study~1 confirms that the prevalence effect not only influences individual judgments but also scales to distort group-level decisions. Though the WoC effect often mitigates individual error, extreme prevalence conditions (e.g., 10\% and 90\%) can reverse this trend. These findings suggest that systematic biases, like the prevalence effect, violate the independence assumptions critical for successful aggregation. Notably, increasing crowd size improves accuracy only when individual-level accuracy is above chance, emphasizing the limitations of naïve aggregation in highly biased environments.

\section{Study 2}
\label{sec:study2}
Study~2 builds on the work of Study~1 by investigating two strategies to mitigate the WoC prevalence effect. One strategy attempts to debias individual-level prevalence effects at the time of judgment using balanced feedback throughout the task. The idea is that if the prevalence effect can be eliminated from the individual-level responses, then it will never have a chance to spread to the crowd level. The second strategy attempts to correct both individual and crowd-level judgments post hoc using recalibration methods.

We test both strategies in a field experiment, using the same task of classifying white blood cells, by collecting annotations through DiagnosUs \citep{press2021centaur}, a crowdsourcing platform specializing in medical and scientific data annotation.
DiagnosUs presents labeling tasks to annotators in the form of contests and users can choose whether to participate in each contest.
In a contest, GS images (images with ground truth labels) are randomly shuffled in with unlabeled images to provide feedback to users and continually assess their performance.

\subsection{Behavioral Methods}

\subsubsection{Participants.}
On DiagnosUs, participants compete against one another to win prize money for each contest.
The amount of money they win depends on their place on the contest leaderboard, which is determined by their performance in the contest.
In order to be eligible to be placed on the leaderboard for a given contest, each participant was required to complete at least 200 trials.
Responses from users who completed fewer than 200 trials were excluded from further analysis.
In each contest across conditions, first place received a \$30 prize, second place received a \$25 prize, third place received a \$20 prize, fourth and fifth place received a \$15 prize, sixth place received a \$10 prize, seventh place received a \$5 prize, and eighth through twelfth place received a \$1 prize.
All other participants did not receive any prize money.

A total of 290 participants were recruited through DiagnosUs for Study~2.
Each condition was presented to users as contests and the app user base was randomly partitioned into four equally sized groups such that users in a given group could only participate in contests from one condition.
As a result, no user competed in contests from more than one condition.
There were 75 participants in the binary choice (BC) 20\% GS blast prevalence condition, 67 participants in the BC 50\% GS blast prevalence condition, 75 participants in the elicited beliefs (EB) 20\% GS blast prevalence condition, and 73 participants in the EB 50\% GS blast prevalence condition.

\subsubsection{Materials.}
A total of 532 images of Wright-stained white blood cells were used.
The images were randomly divided into a QA set containing 300 unique images (150 blast and 150 non-blast) and a GS set containing 232 unique images (116 blast and 116 non-blast).
In order to create the 20\% QA prevalence rate in all conditions, each of the 150 non-blast images was rotated three times (90, 180, and 270 degrees) such that the QA set contained 750 images (150 blast and 600 non-blast) when including rotations. Note that all images were square and do not have a specific orientation (there is no natural ``up'' or ``down'' in cell images).
To create the 20\% GS prevalence rate in the BC 20\% GS blast prevalence and EB 20\% GS blast prevalence conditions, each of the 116 non-blast images was rotated three times such that the GS set contained 580 images (116 blast and 464 non-blast) when including rotations.
No images in the GS set were rotated for the BC 50\% GS blast prevalence and EB 50\% GS blast prevalence conditions since the GS set was already balanced.

\subsubsection{Procedure.}
In all conditions, participants first read instructions explaining that they would be presented with images of white blood cells.
In the EB conditions, participants were informed that their task would be to judge the likelihood that each image contained a blast cell.
In the BC conditions, participants were informed that their task would be to determine whether an image contained a blast cell.
After the initial instructions, participants completed eight practice trials to familiarize themselves with the task.
On each practice trial in the EB conditions, participants were asked ``What is the likelihood that this is a blast cell?''.
Participants entered their response using a slider mechanism and were given feedback in terms of the correct answer and a score which was computed using the binarized, quadratic scoring rule.
Similarly, on each practice trial in the BC conditions, participants were asked ``Do you think this image is a blast cell?''.
Participants would either select \textit{yes} or \textit{no} and they were provided categorical feedback (i.e., the correct answer) after each response.
Images in the practice trials were randomly sampled from the GS set with replacement.

After the eight practice trials, participants could proceed to the competition trials.
The answer prompts and response elicitation mechanisms in the competition trials were identical to those employed in the practice trials in each condition.
For both practice and competition trials, there was no time limit set for each trial so participants could take as long as they liked to make their judgment.
Images in the competition trials were randomly sampled such that one third of the images came from the GS set and two thirds of the images came from the QA set, with the constraint that no user saw the same image twice within the same contest (excluding rotations of the same image).
Participants received feedback on competition trials in the same format as the practice trials, but only on trials where images were sampled from the GS set.
No feedback was provided for responses on QA images.
In the EB condition, participants' performance (and therefore their place on the leaderboard) was based on their average score on competition trials for images in the GS set.
Similarly, in the BC condition, participants' performance was based on their accuracy on competition trials for images in the GS set.

\subsubsection{Recalibration Approach.}

In the EB condition, both participants' responses and WoC labels were recalibrated using the LLO function.
This function transforms the raw probabilities $(p)$ into ones that are more aligned with objective truth \citep{birnbaum1996violations,tversky1995weighing,turner2014forecast}:
\begin{equation}\label{eq:llo}
\ln\!\biggl(\frac{f(p)}{1-f(p)}\biggr) = \alpha \ln\!\biggl(\frac{p}{1-p}\biggr) + \beta.
\end{equation}
A holdout calibration set where we have both probability judgments and ground truth labels is required to estimate the two free parameters, $\alpha$ and $\beta$, of the LLO function.
GS images serve as a natural calibration set because there are existing ground truth labels for these images. Thus, we are able to leverage the existing task structure on DiagnosUs (interleaved GS and QA items) for our recalibration approach.

For individual recalibration, the two free parameters are estimated via maximum likelihood using a participant's responses for images in the GS set.
After the two free parameters are fit using a participant's responses for images in the GS set, the LLO function can then be used to transform all of the participant's responses, both for images in the GS set and QA set.
Note that individual calibration was carried out independently for each participant because there can be large individual differences in the biases that lead to miscalibration across annotators \citep{baron2014two}, so the $\alpha$ and $\beta$ parameters of the LLO function vary across participants.
We will refer to participants' recalibrated subjective probability judgments as rEB (recalibrated elicited belief) judgments.

\subsubsection{WoC Label Generation.}
For BC datasets, the WoC label for a given image was generated by computing the proportion of ``blast'' judgments from nine randomly sampled individual judgments for that image (e.g., if four blast judgments were sampled and five non-blast judgments were sampled, the WoC label would be 44.4\% blast).
For EB and rEB datasets, the WoC label for a given image was generated by taking the mean of nine randomly sampled judgments for that image.
For crowd recalibration of the rEB dataset, the LLO function was fit to the WoC labels for images in the GS set and then used to transform the crowd labels for images in the QA set.
Similar to how individual recalibration was carried out independently for each participant, crowd recalibration was carried out independently for each rEB dataset.
We will refer to rEB datasets with crowd recalibration as rEB w/ CR datasets and rEB datasets without crowd recalibration as rEB w/o CR datasets.

This process of randomly sampling nine judgments and taking the mean to obtain the WoC label for a given image was repeated 100 times for each image to generate 100 BC, EB, rEB w/o CR, and rEB w/ CR WoC datasets for each GS prevalence rate condition, resulting in a total of eight different types of labeled datasets.
We will refer to these different types of labeled datasets as data variants.

\subsubsection{Machine Learning Methods.}
\label{sec:ml}
The 750 images (including all rotations) in the QA set were used for hyperparameter selection and model training/testing.
The images in the QA set were divided into training and testing sets using stratified five-fold cross-validation.
As a result, each model is trained using roughly 80\% of the images and evaluated using roughly 20\% of the images, with the constraints that the proportion of blast images in the training and testing sets is approximately 20\%, each image appears in the testing set an equal number of times across all folds, and all rotations of a given image are included within the same set (either training or testing) on each fold.
Each CNN was trained via transfer learning using PyTorch \citep{paszke2019pytorch}.
To implement this, we modified a GoogLeNet CNN \citep{szegedy2015going} which was pre-trained on the ImageNet database using the same transfer learning procedure in \cite{holmes2020joint}.
Each CNN was trained to minimize the cross-entropy loss between the CNN output and the training labels using the Adam optimizer.

To identify the optimal hyperparameters for models trained on each type of labeled dataset, we performed a grid search over the number of training epochs (5, 10, 15, 20, and 25), learning rate ($10^{-4}$, $5\times10^{-4}$, and $10^{-3}$), L2-regularization strength ($10^{-4}$, $5\times10^{-4}$, $10^{-3}$, $5\times10^{-3}$, and $10^{-2}$), and minibatch size (16, 24, and 32).
For hyperparameter selection, the stratified five-fold cross-validation process was repeated six times resulting in 30 unique sets of training and testing splits.
The hyperparameters of the model that yielded the lowest cross-entropy loss between the model's output and the crowd labels on the testing split averaged across the 30 testing splits were deemed optimal.
After the optimal hyperparameters were identified for each type of labeled dataset (so there were eight unique sets of optimal hyperparameters), the model training and testing process was again carried out using these hyperparameters with more training/testing splits.
For the CNN training and evaluation, the stratified five-fold cross-validation process was repeated 20 times resulting in 100 unique sets of training and testing splits.
The models were still trained using the crowd labels, but now were evaluated with respect to the true labels on out-of-sample images.

\subsection{Behavioral Results}

\subsubsection{Individual Results.}

\begin{figure}
\FIGURE
{\includegraphics[width=\linewidth]{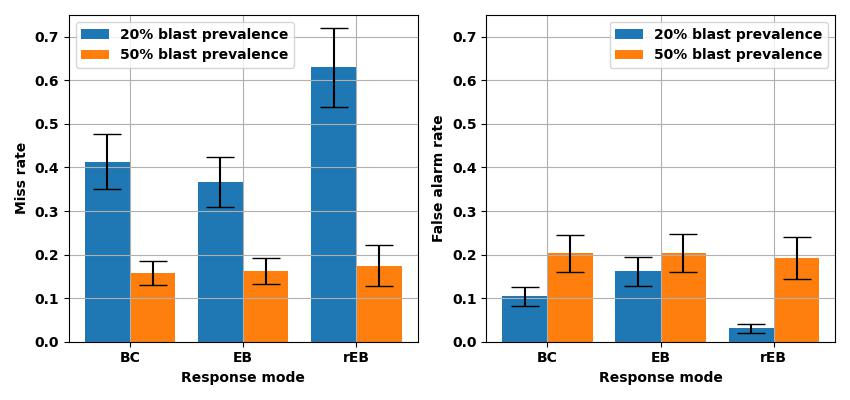}}
{Individual-level miss rates and false alarm rates across the various conditions and response modes in Study~2. Blue bars show the conditions where the GS prevalence rate matched that of the QA images (i.e., 20\%). Orange bars show the conditions where the GS prevalence rate was 50\%. Errorbars represent the 95\% confidence intervals for the mean miss rate and false alarm rate across the participants in each condition and response mode.\label{fig:study2-indiv-mr-far}}
{}
\end{figure}
Across all response modes, the miss rate decreases and the false alarm rate increases when the GS blast prevalence rate increases from 20\% to 50\%, as shown in Figure~\ref{fig:study2-indiv-mr-far}.
Overall, using balanced feedback is effective at counteracting the prevalence effect on the QA set because the errors are much more equally distributed across misses and false alarms compared to when participants are receiving unbalanced feedback.
Recalibrating individual responses, at least in the 20\% GS blast prevalence condition, appears to make the prevalence effect more extreme.
That is, the individual miss rate with the BC and EB response modes is around 35--40\%, whereas in the rEB response mode it is around 60\%; likewise, the individual false alarm rate is around 10--15\% in the BC and EB response modes, but only around 3\% in the rEB response mode.
In the 50\% GS blast prevalence condition, the miss rates and false alarm rates do not appear to vary much across the BC, EB, or rEB datasets.

\subsubsection{Crowd Results.}
\begin{figure}
\FIGURE
{\includegraphics[width=\linewidth]{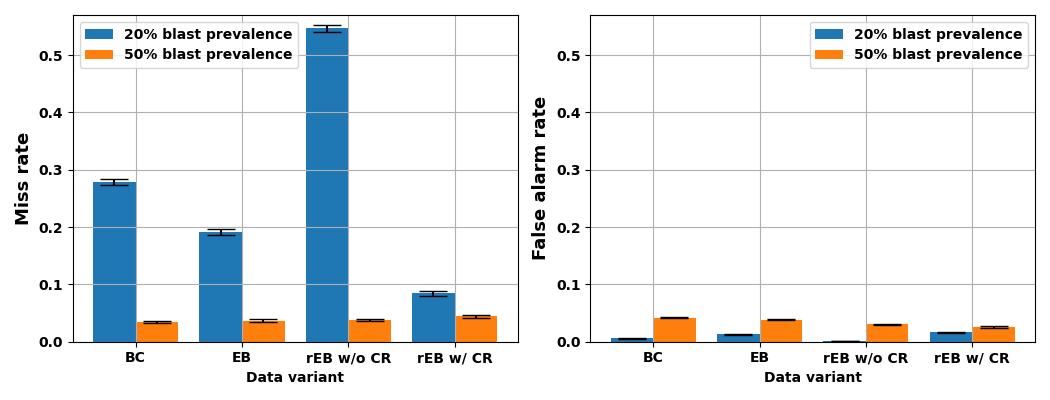}}
{Crowd-level miss rates and false alarm rates for the eight data variants in Study~2. Errorbars represent the 95\% confidence intervals for the mean miss rate and false alarm rate across the 100 sets of labels for each type of labeled dataset.\label{fig:study2-crowd-mr-far}}
{}
\end{figure}

The crowd results shown in Figure~\ref{fig:study2-crowd-mr-far} appear to diverge from the individual-level results in some aspects.
The commonalities are that miss rate always decreases and false alarm rate always increases when GS blast prevalence rate increases from 20\% to 50\%, and the errors are much more evenly balanced in the balanced feedback conditions compared to the unbalanced feedback conditions.

Crowd recalibration improves the miss rate of the rEB datasets in the 20\% GS blast prevalence feedback condition.
Now, rEB w/ CR produces the lowest miss rate, around 9\%, compared to the BC, EB, and rEB w/o CR datasets in the 20\% GS blast prevalence condition.
The improved miss rate does come at the cost of a higher false alarm rate compared to the BC, EB, and rEB w/o CR datasets, but the false alarm rate for the rEB w/ CR datasets is still only around 3\%.
Also, unlike the individual results, the EB response mode appears to produce a much lower miss rate compared to the BC response mode in the 20\% prevalence condition.
This suggests that the uncertainty conveyed in subjective probability judgments can be used to reduce the prevalence effect even without recalibrating the probabilities.

\begin{figure}
\FIGURE
{\includegraphics[width=0.9\linewidth]{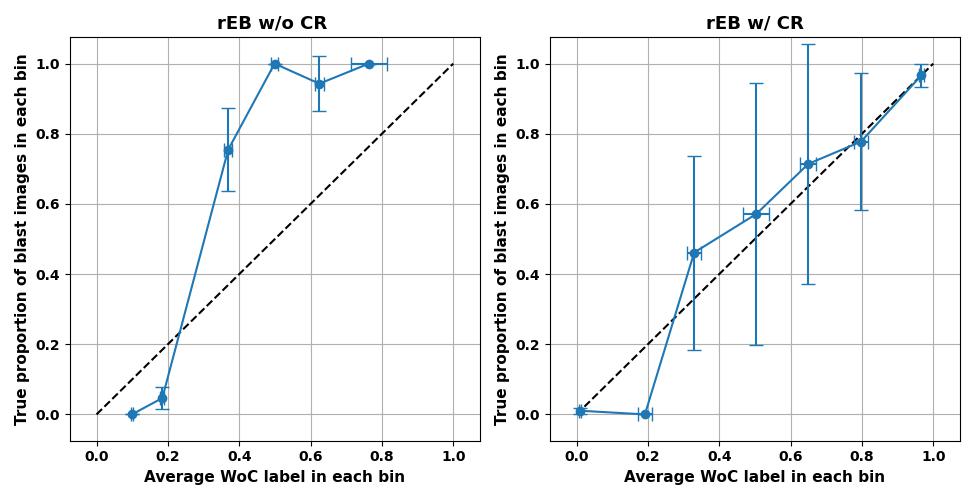}}
{Calibration curves for the two example rEB w/o CR and rEB w/ CR datasets in the 20\% GS prevalence condition.
The x-errorbars represent the 95\% confidence interval around the mean WoC label in each bin and y-errorbars represent the 95\% confidence interval around the true proportion of blast images in each bin.\label{fig:rwoc-impact}}
{}
\end{figure}

The improved performance of the rEB w/ CR datasets in the 20\% GS blast prevalence condition arises from recalibration at the crowd level.
To help illustrate why crowd recalibration improves the miss rate in the 20\% GS blast prevalence condition, Figure~\ref{fig:rwoc-impact} depicts the calibration curves for example rEB w/o CR and rEB w/ CR datasets.
Calibration curves are generated by first binning the WoC labels into seven equally spaced bins.
We used seven bins because seven bins most clearly illustrate the impact of crowd recalibration on the calibration curves of the rEB datasets.
Each point on the calibration curve corresponds to one bin.
The x-value of each point is equal to the average WoC label in the corresponding bin and the y-value of each point is equal to the true proportion of blast images in each bin.
For example, for the rEB w/o CR dataset, the coordinates of the leftmost point in the left bin indicate that, for all labels between $[0, 1/7)$, the average WoC label is equal to 0.1 and the true proportion of blast images in this bin is 0.
If all points lie on the $y = x$ line, then the WoC labels are perfectly calibrated, in the sense that the average WoC label is equal to the true proportion of blast cells in each bin.
The further the calibration curve deviates from the $y = x$ line, the poorer the calibration of the dataset.

Based on Figure~\ref{fig:rwoc-impact}, the rEB w/o CR dataset systematically underestimates the likelihood that the images are blast cells.
For example, the third point in the plot indicates that roughly 75\% of the images with WoC labels in $[2/7, 3/7)$ were blast images that were classified as non-blast.
Crowd recalibration mitigates this bias by transforming the WoC labels such that they are shifted upward towards higher values.
As a result, many of the images that were incorrectly classified as non-blast (misses) are now correctly classified as blast, as their WoC labels were shifted from $<0.5$ to $>0.5$.
By shifting many of the WoC labels on blast images from $<0.5$ to $>0.5$, crowd recalibration results in a large decrease in miss rate without having a major impact on false alarm rate, as seen in Figure~\ref{fig:study2-crowd-mr-far}.

The improvements in calibration are quantified using expected calibration error (ECE) \citep{guo2017calibration}.
ECE is computed by first dividing the crowd labels for a single dataset into $N$ equally spaced bins (here we used $N = 10$).
The calibration error for a given bin is equal to the distance between the average label in that bin and the proportion of blast cell images in that bin.
ECE is then defined as the weighted sum of the calibration errors across bins, where the weight for a given bin is equal to the number of images in that bin divided by the total number of images in the dataset (750).

\begin{figure}
\FIGURE
{\includegraphics[width=.8\linewidth]{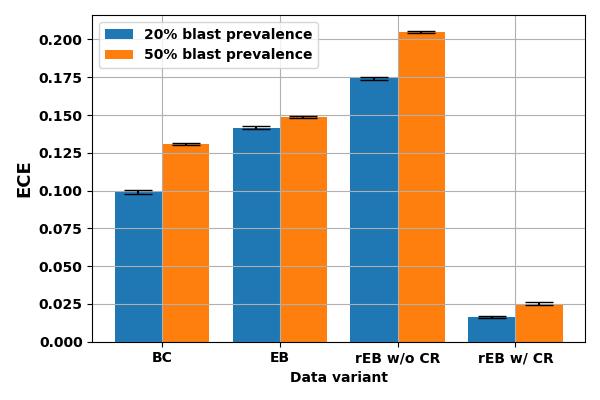}}
{ECE for the eight types of data variants in Study~2. Errorbars represent the 95\% confidence intervals for the mean ECE across the 100 sets of labels for each type of labeled dataset.\label{fig:study2-crowd-ece}}
{}
\end{figure}

In both prevalence conditions, the rEB w/ CR datasets produced the best calibrated datasets, as shown in Figure~\ref{fig:study2-crowd-ece}.
This results from the fact that the rEB w/ CR datasets were recalibrated at the crowd level, whereas the BC, EB, and rEB w/o CR datasets were not calibrated at the crowd-level.
In contrast, the rEB w/o CR datasets produced the worst calibrated datasets.

\begin{figure}
\FIGURE
{\includegraphics[width=\linewidth]{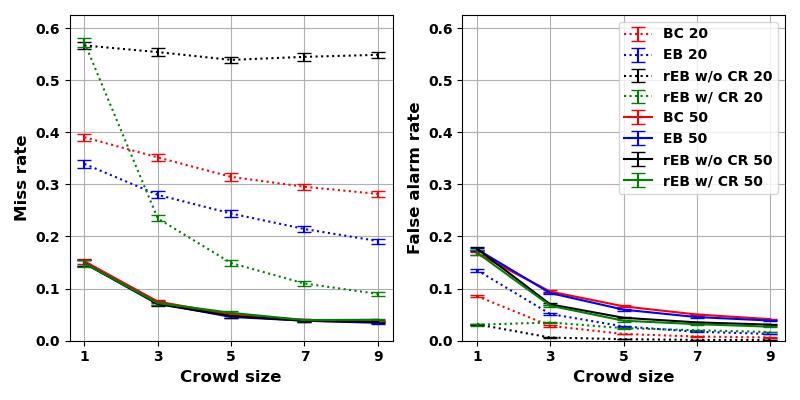}}
{Crowd-level miss rate and false alarm rate as a function of the number of judgments per image ($N$) in Study~2.
The errorbars represent the 95\% confidence interval for the mean miss rate and false alarm rate across the 100 simulations for each value of $N$.\label{fig:study2-crowd-size}}
{}
\end{figure}

In Study~1, crowd-level miss rate in the 10\% prevalence condition increased with crowd size because the average individual-level miss rate was greater than 0.5.
Hence, we might expect to see the same in Study~2 for the rEB datasets in the 20\% GS blast prevalence condition because the individual-level rEB miss rate for this condition is 0.58 in Study~2.
However, Figure~\ref{fig:study2-crowd-size} shows that crowd-level miss rate does not increase monotonically with crowd size for either the rEB w/o CR or rEB w/ CR datasets.
For the rEB w/o CR dataset, the miss rate flutters around 0.55 and does not increase or decrease monotonically.
For the rEB w/ CR dataset, the miss rate sharply decreases monotonically.
Crowd recalibration produces this result because it corrects for the systematic underestimation bias illustrated in Figure~\ref{fig:rwoc-impact}.
As the crowd size increases, the quality of the recalibration improves, resulting in a decreasing miss rate as crowd size increases for the rEB w/ CR datasets.

There does appear to be some minor tradeoff though, as the false alarm rate increases when crowd size increases from one to three votes per image for the rEB w/ CR datasets in the 20\% GS prevalence condition, before decreasing with the larger crowd sizes.
This shows that even with crowd recalibration, a small crowd size can occasionally produce negative (albeit extremely minor) effects on miss rate or false alarm rate.
Otherwise, the impact of crowd size on miss rate and false alarm rate is similar to that of Study~1, with both error rates decreasing as the crowd size increases.

\subsection{Machine Learning Results}
\begin{figure}
\FIGURE
{\includegraphics[width=\linewidth]{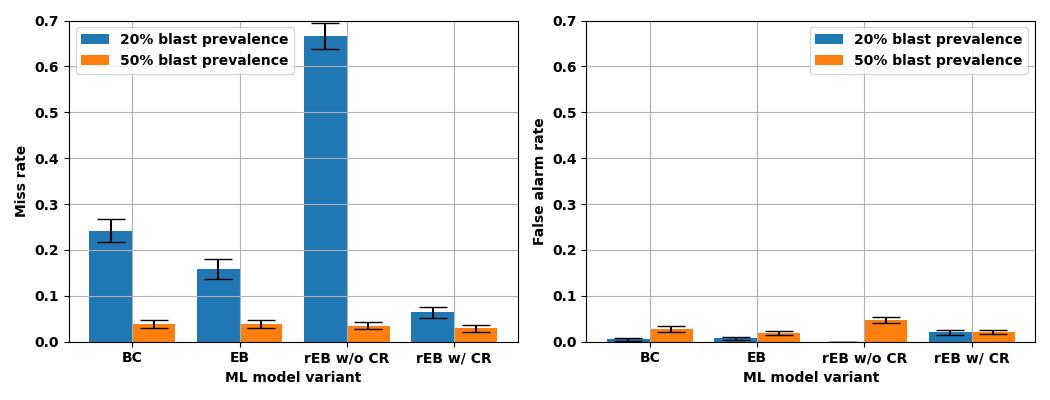}}
{Test miss rates and false alarm rates for the ML models trained on the eight types of data variants. Errorbars represent the 95\% confidence intervals for the mean miss rate and mean false alarm rate across the 100 models for each type of labeled dataset.\label{fig:study2-ml-mr-far}}
{}
\end{figure}

The miss rate, false alarm rate, and calibration of the ML models were measured in the same way as they were measured for the data variants, except that the ML models were only evaluated using their output on the testing set.
Figure~\ref{fig:study2-ml-mr-far} contains the miss rate and false alarm rate results for all eight ML model variants.

The ML results follow very similarly to the crowd-level results.
That is, the miss rate across all response modes is significantly higher and the false alarm rate is slightly lower in the 20\% GS blast prevalence conditions.
Also, the miss rate and false alarm rates are much more balanced in the 50\% GS prevalence conditions compared to the 20\% GS blast prevalence conditions.
Additionally, the miss rate for the models trained on the EB datasets from the 20\% GS blast prevalence conditions is significantly lower compared to the models trained on the BC datasets from the 20\% GS blast prevalence condition. Lastly, in the 20\% GS blast prevalence condition, models trained on the rEB w/ CR datasets had the lowest miss rate.

The ML models do appear to produce slightly lower mean miss rates compared to that of the datasets they are trained on in the 20\% GS blast prevalence conditions, except for the models trained on the rEB w/o CR datasets.
For example, the mean miss rate for the models trained on the BC datasets is equal to 0.24, whereas the mean miss rate for BC datasets themselves is equal to 0.28.
This suggests that the models may not fully inherit the biases embedded in the data variants that lead to the high miss rates in the 20\% GS blast prevalence conditions.

\begin{figure}
\FIGURE
{\includegraphics[width=0.6\linewidth]{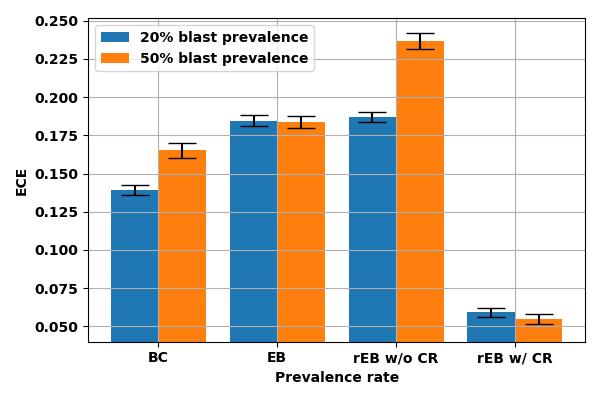}}
{Expected calibration error (ECE) of the ML models trained on the various data variants in Study~2.\label{fig:study2-ml-ece}}
{}
\end{figure}
Based on Figure~\ref{fig:study2-ml-ece}, the ECE of the models also follows similar trends as those seen in the labeled datasets.
The ECE of the models trained on rEB w/ CR datasets is far lower compared to that of models trained on the BC, EB, and rEB w/o CR datasets, with the rEB w/o CR datasets having the highest ECE.
However, contrary to the mean miss rate and false alarm rate results, the ECE of the ML models are worse than the datasets they are trained on across the board.

\subsection{Discussion}
Study~2 investigates approaches to mitigate the WoC prevalence effect documented in Study~1. In a field experiment conducted in a crowdsourcing context, we demonstrate that both balanced feedback and recalibration help mitigate the prevalence effect at the crowd-level and in the ML models trained on the crowd-level data. Most notably, under conditions where we would expect aggregation to exacerbate bias as the crowd size increases, crowd-level recalibration corrects this bias such that the improvements typically afforded by WoC are retained.

\section{General Discussion}

\label{sec:discussion}

This paper studies an operations problem that sits at the center of the AI lifecycle: how to design human labeling pipelines for \emph{rare-event} classification so that aggregated labels, and the ML models trained on those labels, remain accurate and reliable.
Across two studies in medical image annotation, we show that rare-event environments induce the prevalence effect, which inflates misses (false negatives) and generates positively correlated errors across annotators \citep{wolfe2005rare,wolfe2007low,gur2003prevalence,horowitz2017prevalence,trueblood2021disentangling}.
Because the value of redundancy and aggregation depends on individual errors not being strongly aligned \citep{hong2004groups,davis2014crowd}, these systematic cognitive biases create an operational failure: collecting more labels and applying naive WoC rules can deliver diminishing returns and, in extreme regimes, can even amplify the miss-heavy error profile.

Our results also show that this is a \emph{design problem}, not a fixed property of annotators.
In DiagnosUs contests \citep{press2021centaur}, the platform controls how many and what kind of gold-standard (GS) items appear.
Holding the prevalence of positives in the unlabeled QA stream fixed (20\% blasts), we demonstrate that changing the prevalence in the GS feedback stream materially changes behavior and downstream outcomes.
Providing \emph{balanced} GS feedback (50\% blasts) yields a more balanced miss/false-alarm tradeoff at both the individual and crowd levels.
Further, moving from binary labels to elicited probabilities provides richer information that improves aggregation in low-prevalence conditions.
Finally, when probabilistic judgments are available, a scalable \emph{pipeline-level} recalibration approach, implemented here via a linear-in-log-odds (LLO) transformation \citep{birnbaum1996violations,turner2014forecast,han2022recalibrating}, can correct systematic underestimation of the rare class that persists after naive aggregation.

The operational relevance of these interventions is underscored by their downstream impact on ML.
The miss/false-alarm patterns observed in the crowd labels largely carry through to convolutional neural networks trained on those labels, implying that annotation design choices propagate into model behavior.
In our lowest-prevalence feedback regime, crowd-level recalibration of aggregated probabilities reduces the crowd miss rate to roughly 9\% while keeping false alarms near 3\% (Figure~\ref{fig:study2-crowd-mr-far}); models trained on these recalibrated labels exhibit corresponding improvements in rare-event detection (Figure~\ref{fig:study2-ml-mr-far}).
Recalibrated labels also substantially improve probabilistic reliability, as measured by expected calibration error (ECE) \citep{guo2017calibration}, relative to alternative label variants (Figures~\ref{fig:study2-crowd-ece} and \ref{fig:study2-ml-ece}).
That said, model training can degrade calibration relative to the training labels, suggesting that high-quality data operations and model-side calibration are complements rather than substitutes.

\subsection{Implications for managing AI data operations}

Taken together, the findings provide several actionable lessons for organizations that rely on large-scale human labeling for rare-event AI systems.

\paragraph{Treat the GS feedback stream as a policy lever.}
In many annotation operations, GS items are used simultaneously for training and assessing annotator performance.
Our field evidence shows that the \emph{composition} of this stream is not innocuous: it changes the experienced base rate and, consequently, how workers make decisions.
A practical implication is to decouple (i) a \emph{monitoring} set whose prevalence matches production for unbiased measurement from (ii) a \emph{feedback/training} stream whose prevalence is chosen to manage known cognitive biases.
Balanced or partially balanced feedback can be especially valuable when the operational cost of misses dominates.

\paragraph{Elicit probabilities when feasible and score them properly.}
Binary labels discard information about uncertainty.
In our setting, elicited beliefs (paired with a proper scoring rule in feedback) improve aggregation under class imbalance by allowing the pipeline to weight marginal cases differently than confident negatives.
More broadly, probabilistic interfaces create the input required for scalable post-processing (recalibration), which can improve downstream AI.

\paragraph{Correct bias at the pipeline level, not only at the worker level.}
Individual-level recalibration can be unstable when each worker sees only a modest number of GS trials, and it can overcorrect in low-prevalence environments.
By contrast, crowd-level recalibration uses the pooled GS information embedded in the pipeline, yielding a more stable correction that improves both error rates and calibration.
Operationally, this suggests viewing recalibration as a standard quality-control method that can be updated continuously as more GS feedback accumulates.

\paragraph{Evaluate annotation operations with operationally-aligned metrics.}
Unconditional accuracy can mask miss-heavy failures in rare-event tasks.
Our results illustrate the importance of tracking misses, false alarms, and calibration jointly, and of validating interventions not only on label quality but also on the behavior of models trained on those labels.

\subsection{Limitations and directions for future research}

This work has several limitations that point to fruitful directions for future research.
First, our empirical setting focuses on a single binary medical image task, and the strength and form of prevalence-induced biases may differ in other domains (e.g., security screening or content moderation) and in multi-class labeling environments \citep{wolfe2013prevalence,chandrasekharan2018internet}.
Second, DiagnosUs contests combine incentives, feedback, and selection (workers opt into contests), which mirrors many real operations but may interact with prevalence in ways that deserve further study (e.g., workforce composition and learning dynamics).
Third, we manipulate GS prevalence at two levels (20\% vs.\ 50\%); characterizing optimal feedback policies under alternative miss/false-alarm costs, budget constraints, and prevalence trajectories is an important operational design problem.
Finally, our downstream analysis shows that model fitting can partially erode calibration, highlighting an opportunity for integrated pipeline design that jointly selects feedback policies, aggregation rules, and model-side calibration methods.

Overall, the results emphasize a core message for AI in operations: reliable ML systems require not only better models, but also better \emph{data operations}.
When rare events distort human judgment, redundancy alone is insufficient; instead, organizations must actively manage the feedback, elicitation, and calibration components of the labeling process so that both labels and models remain fit for operational use.

\section{Conclusion}

\label{sec:conclusion}

Operational AI systems increasingly depend on human-in-the-loop pipelines to label rare but consequential events.
This paper shows that class imbalance can create systematic, shared cognitive biases that inflate misses and limit the returns to redundancy and naive aggregation.
Using evidence from prior experiments and a field experiment on a live medical crowdsourcing platform, we identify three scalable levers that organizations can deploy: (i) designing the prevalence of positives in the gold-standard feedback stream, (ii) eliciting subjective probabilities rather than only binary labels, and (iii) applying pipeline-level recalibration to correct systematic bias.
Together, these levers restore the benefits of aggregation in rare-event environments and improve the reliability of ML models trained on the resulting data.

More broadly, the findings reinforce an operations perspective on AI: model performance and reliability are jointly determined by algorithms \emph{and} the end-to-end processes that generate, validate, and maintain data.
For AI applications where misses are especially costly, managing the annotation operation, including how feedback is delivered and how probabilistic information is post-processed, is a central determinant of system performance in deployment.

\bibliographystyle{informs2014}
\bibliography{citations_msom}

\end{document}